\documentclass[12pt]{article}
\usepackage{amsfonts}
\usepackage{bbm}
\usepackage{mathrsfs}
\usepackage{color}
\usepackage{geometry}             
\usepackage{graphicx}
\usepackage{amssymb}
\usepackage{amsmath}
\usepackage{epstopdf}
\usepackage{cite}

\usepackage[hyperindex=true,
          pdfstartview=FitH,
          bookmarksnumbered=true,
          bookmarksopen=true,
          citecolor=blue,
          linkcolor=blue,
          colorlinks=true,
          pdfborder=001,
          unicode]{hyperref}

\allowdisplaybreaks

\begin{document}

\title{Compressible forced viscous fluid from \\
product Einstein manifolds}
\author{Xin Hao\thanks{{\em email}: \href{mailto:shanehowe@mail.nankai.edu.cn}
{shanehowe@mail.nankai.edu.cn}},  
Bin Wu\thanks{{\em email}: \href{mailto:wubin@mail.nankai.edu.cn}
{wubin@mail.nankai.edu.cn}}  
and Liu Zhao\thanks{{\em email}: \href{mailto:lzhao@nankai.edu.cn}
{lzhao@nankai.edu.cn}, correspondence author.}\\
School of Physics, Nankai University, Tianjin 300071, China
}
\date{January 19, 2015}
\maketitle

\begin{abstract}
We consider the fluctuation modes around a hypersurface $\Sigma_c$ in 
a $(d+2)$-dimensional product Einstein manifold, with $\Sigma_c$ taken either 
near the horizon or at some finite cutoff from the horizon. 
By mapping the equations that governs the lowest nontrivial order 
of the fluctuation modes into a system of partial differential equations 
on a flat Newtonian spacetime, a system of compressible, forced viscous
fluid is realized. This result generalizes the non bulk/boundary 
holographic duality constructed by us recently to 
the case of a different background geometry.

\vspace{4mm}

\vspace{4mm}

\end{abstract}

\section{Introduction}

During the last two decades our understanding on the properties of relativistic
gravitation has been strengthened considerably. The major new concept which
centralizes the studies on gravitation is holographic duality
\cite{t'Hooft,susskind} originated from the area law of black hole
entropy and has been realized in a number of different physical configurations,
mostly with gravity in the bulk and some other physical system on the
boundary. Among the various realizations of holographic duality, the most
important ones can be summarized as follows. First comes the well-known AdS/CFT
correspondence \cite{Maldacena:1997re,Gubser:1998bc,Witten:1998qj}, which
has led to important applications and remains an active subject of
study. Generalization of AdS/CFT has resulted in the so-called AdS/QCD 
\cite{Kovtun:2003fk,Brodsky:2010th}
and AdS/CMT correspondences. In particular, 
in the AdS/CMT correspondence, important
progresses have been made in the holographic understanding of phase transitions
such as superconductivity \cite{Hartnoll:2008ht,Hartnoll:2008hz,
Horowitz:2010uu,Herzog:2009cm},  
superfluidity \cite{Herzog:2008eq,Herzog:2009cm} and so on. Another important
aspect of holographic duality is the interpretation of quantum entanglement
entropy using a holographic setup 
\cite{Nishioka:2009ey,Takayanagi:2012wk}. This line of research seems to
be hopeful in interpreting the origin of the area law of the black hole entropy,
see \cite{Solodukhin:2011vya} for a living review. Besides all these, 
there is yet
another important realization of holographic duality, i.e. the Gravity/Fluid
correspondence, see \cite{Bredberg:2010ky,strominger1,Compere,cai1,Eling,
Compere:2012mt,Eling:2011cl,Bai:2012ci,Zou:2013ix,
Hu:2013dza,Niu:2011gu,Cai,Cai:2012mg,1204.2029,
Strominger2,xiaoning,Huang:2011kj,Ying2,Wu:2013kqa,Ling:2013kua,WB1,WB2,
Cai:2014ywa,Cai:2014sua,HWZ} for an incomplete list of related works. 
Unlike other realizations of holographic duality which
match the full bulk gravitational degrees of
freedom to those of the boundary systems, the Gravity/Fluid correspondence
relates only the fluctuation modes of gravity in the bulk to a fluid system on
the boundary, and the fluid is usually incompressible.
Moreover, the sources of the bulk gravitation contributes to the boundary
fluid as an external force, therefore, fluctuation modes of vacuum gravity
in the bulk constitute a force-free fluid system on the boundary.

In spite of the numerous success of the theoretical realizations of
holographic dual and their applications, a mathematically rigorous definition
for holographic duality is still missing. Practically, anything that relates
gravity in the bulk to a system on the boundary can be viewed as a
realization of holographic duality. In the recent work \cite{HWZ}, we have made
one further step which breaks the bulk/boundary picture
of holographic duality. To be more concrete, what we have realized is
a correspondence between fluctuation modes around a vacuum black hole
solution with (possibly) curved horizon to a forced, compressible fluid
with a stationary density distribution in a flat Newtonian spacetime in one
less dimension. In contrast of most well studied cases of Gravity/Fluid
correspondences, the fluid system in our work does not live on the boundary
of the black hole solution. Moreover, the dual fluid is compressible and
subjects to some extra force, although the background gravity solution
is source free. This result indicates that holographic duality can be
understood beyond the bulk/boundary setting. It is certainly of interests to
have more examples of this kind, which may hopefully led to deeper
understandings about the essence of holographic duality.

In this work we shall present some further example along the lines of
\cite{HWZ}. Instead of a black hole background, we consider fluctuations
around a $(d+2)$-dimensional product Einstein manifold $\mathcal{M}
=\mathcal{M}_1 \times \mathcal{M}_2$, where $\mathrm{dim}(\mathcal{M}_1)=2, 
\mathrm{dim}(\mathcal{M}_1)=d$. Following a process which is parallel
to that made in \cite{HWZ}, we will show that similar
fluid dual can be realized in a flat Newtonian spacetime with one less
dimension. Moreover, we will show that the near horizon limit is not essential
in the construction -- we can realize similar flat space fluid system
by looking at the fluctuation modes around a finite cutoff surface
in the background geometry as well. Since the technics used is extremely similar
to that of \cite{HWZ}, we shall be as brief as possible in the main text and
just present the brief route leading to the results.

\section{Product Einstein manifolds in $(d+2)$-dimensions
and hypersurface geometry}

The $(d+2)$ dimensional product Einstein manifold $\mathcal{M}$ 
which we consider is equipped with the line element
\begin{align}
\mathrm{d}s^2_{\mathcal{M}}=g_{\mu\nu} \mathrm{d}x^\mu \mathrm{d}x^\nu
=-f(r) \mathrm{d}u^2 + 2\mathrm{d}u \mathrm{d}r 
+ e^{\Phi (x)} \delta_{ij} \mathrm{d}x^i \mathrm{d}x^j,   \label{bm}
\end{align}
where $(u,r)$ and $(x^i)$ are coordinates on $\mathcal{M}_1$ and $\mathcal{M}_2$
respectively,
\begin{align}
f(r) = 1 -\omega r
- \frac{2 \Lambda}{d} r^2,   \label{bm1}
\end{align}
and $\Phi(x)$ obeys the following system of equations:
\begin{align}
& \delta^{jk} \partial_j \partial_k \Phi
+ \frac{d-2}{2}\bigg( 2 \partial_i^2 \Phi
+ \delta^{jk} \partial_j \Phi \partial_k \Phi
- (\partial_i \Phi)^2  \bigg)
+ \frac{4}{d} \Lambda e^\Phi=0,\nonumber\\
&\qquad \qquad \qquad (\mbox{no summation over } i)
\label{GLveq} \\
& (d-2) \bigg(\partial_i \partial_j \Phi
- \frac{1}{2} \partial_i \Phi \partial_j \Phi\bigg)=0 \qquad (i \neq j).
\label{GLveq2}
\end{align}
For generic $d$, we present a special solution to the equations \eqref{GLveq} 
and \eqref{GLveq2} in the appendix. More solutions might exist because these 
equations are nonlinear. Given any solution to eqs. \eqref{GLveq} 
and \eqref{GLveq2}, the metric $g_{\mu\nu}$ of the total manifold 
$\mathcal{M}$ obeys the vacuum Einstein equation
\begin{align}
G_{\mu\nu} = -\Lambda g_{\mu\nu}. \label{eins}
\end{align}
We assume that $\omega>0$ and is sufficiently large when $\Lambda<0$, so
that $f(r)$ always has zeros. Let the (biggest, if $\Lambda<0$) zero of $f(r)$
be $r_h$, which represents the radius of a horizon.
Consider a $(d+1)$-dimensional timelike hypersurface $\Sigma_c$
located at $r=r_c$. Evidently, we need $r_c>r_h$ when $\Lambda<0$ and
$r_c<r_h$ when $\Lambda\geq 0$ in order to ensure that $\Sigma_c$ is timelike.
The induced line element on this hypersurface is given by
\begin{align}
 \mathrm{d}s^2_{\Sigma_c}=\gamma_{ab}\mathrm{d}x^a\mathrm{d}x^b
 =& -f(r_c)\mathrm{d}u^2 + e^\Phi \delta_{ij}
 \mathrm{d}x^i \mathrm{d}x^j       \nonumber
 \\=& -(\mathrm{d}x^0)^2
 + e^\Phi \delta_{ij} \mathrm{d}x^i \mathrm{d}x^j     \nonumber
\\=& -\frac{1}{\lambda^2} \mathrm{d}\tau^2
+ e^\Phi \delta_{ij} \mathrm{d}x^i \mathrm{d}x^j, \label{dsd}
\end{align}
where $\tau =  (\lambda \sqrt{f_c})\, u$, $x^0 =\sqrt{f_c}\, u$,
$f_c$ is a shorthand for $f(r_c)$. Similar notations have been used in
\cite{HWZ}.
One can easily promote the hypersurface tensor $\gamma_{ab}$ to a bulk
tensor $\gamma_{\mu\nu}$ and define the extrinsic curvature of $\Sigma_c$ as
$K_{\mu\nu} = \frac{1}{2} \mathscr{L}_n \gamma_{\mu\nu}$,
with $n^\mu$ being a unit vector field which is normal to $\Sigma_c$ at
$r=r_c$. In the coordinates $(u,r,x^i)$, $n^\mu$ has components
\[
n^\mu = \bigg(\frac{1}{\sqrt{f}},\sqrt{f},0,\cdots,0\bigg).
\]

The projection of \eqref{eins} on $\Sigma_c$ yields the
momentum and Hamiltonian constraints,
\begin{align}
&D_a (K^a_{\ b} - \gamma^a_{\ b} K) = 0  \label{momentum c},\\
&\hat{R} + K^{ab} K_{ab} - K^2 = 2 \Lambda,
\end{align}
where $\hat{R}$ is the Ricci scalar of $\Sigma_c$, $D_a$ is the covariant
derivative that is compatible with $\gamma_{ab}$. In terms of the Brown-York
tensor $t_{ab}=\gamma_{ab}K-K_{ab}$, the momentum and Hamiltonian constraints can be rewritten as
\begin{align}
&D_a t^a_{\ b} = 0,       \label{momc}\\
&\hat{R} + t^a_{\ b} t^b_{\ a} - \frac{t^2}{d}
= 2 \Lambda.    \label{hamc}
\end{align}
Explicit values of the extrinsic curvature and the Brown-York tensor
will be made use of in the following context, so we present these values
by direct calculations,
\begin{align}
& K^{\tau}_{\ \tau} = \frac{f'_c}{2\sqrt{f_c}}, && K^{\tau}_{\ i}=0, \nonumber
\\
& K^i_{\ j} = 0, && K = K^a_{\ a}= \frac{f'_c}{2\sqrt{f_c}}.
\end{align}
These in turn lead to the background Brown-York tensor
\begin{align}
& t^{\tau(B)}_{\ \tau} = 0,    && t^{\tau(B)}_{\ i} = 0,    \nonumber\\
& t^{i(B)}_{\ j} =\frac{ f'_c }{2\sqrt{f_c}}
\delta^i_{\ j}, && t^{(B)} =t^{a(B)}_{\ a}
= \frac{d}{2}\frac{f'_c}{\sqrt{f_c}}. \label{tb}
\end{align}
We have intentionally added a superscript $(B)$ to indicate that these are the
values with respect to the background geometry. When considering
fluctuation modes, these values must be supplemented with fluctuation
modifications.

\section{Petrov I boundary condition and near horizon limit}

In this section we shall consider the case with $\Sigma_c$ placed very closed to
the horizon, i.e. $r_c-r_h={{\epsilon}}\,\alpha^2\lambda^2$, where
$\epsilon= 1$ if $\Lambda < 0$, $\epsilon= -1$ if $\Lambda \geq 0$,
and $\lambda \to 0$ is the same parameter appeared in \eqref{dsd}.
The positive constant $\alpha$ is introduced to eliminate some un-necessary
complexity when taking the near horizon limit.
We will focus ourselves on the fluctuation
modes around $\Sigma_c$ and pay particular attention to the Petrov I boundary
condition
\begin{align}
C_{(l)i(l)j}|_{\Sigma_c}=l^\mu(m_i )^\nu l^\rho(m_j)^\sigma
C_{\mu\nu\rho\sigma}|_{\Sigma_c}=0,
\end{align}
where $C_{\mu\nu\rho\sigma}$ represents the bulk Weyl tensor, and
\begin{align}
&l^\mu
=\frac{1}{\sqrt{2}}\bigg(\frac{1}{\sqrt f}
(\partial_u)^\mu-n^\mu\bigg),\quad
k^\mu =\frac{1}{\sqrt{2}}\bigg(\frac{1}{\sqrt f}
(\partial_u)^\mu+n^\mu\bigg),\nonumber\\
&(m_i)^\mu = e^{-\frac{1}{2}\Phi} (\partial_i)^\mu
\end{align}
constitute a set of Newman-Penrose basis vector field which obeys
\begin{align}
&  l^2=k^2=0,\ ,(k,l)=1,\ ,(l,m_{i})=(k,m_{i})=0,\ ,(m_{i},m_{j})=\delta_{ij}.
\end{align}
When restricted on $\Sigma_c$, we can write
\begin{align}
&l^\mu|_{\Sigma_c}
=\frac{1}{\sqrt{2}}\big((\partial_0)^\mu-n^\mu\big),\quad
k^\mu|_{\Sigma_c} =\frac{1}{\sqrt{2}}\big((\partial_0)^\mu+n^\mu\big).
\end{align}
Therefore the boundary condition can also be cast in the form
\begin{align}
C_{0i0j}+C_{0ij(n)}+C_{0ji(n)}+C_{i(n)j(n)}=0,       \label{pjbc}
\end{align}
with
\begin{align}
&C_{abcd}
= \gamma^\mu_{\ a} \gamma^\nu_{\ b} \gamma^\sigma_{\ c} \gamma^\rho_{\ d}
C_{\mu\nu\sigma\rho}
= \hat{R}_{abcd} + K_{ad} K_{bc} - K_{ac} K_{bd}
-\frac{{4}\Lambda}{d(d+1)} \gamma_{a[c} \gamma_{d]b},        \nonumber
\\
& C_{abc(n)} =
\gamma^\mu_{\ a} \gamma^\nu_{\ b} \gamma^\sigma_{\ c} n^\rho
C_{\mu\nu\sigma\rho}
= D_a K_{bc} - D_b K_{ac},            \nonumber
\\
& C_{a(n)b(n)}
= \gamma^\mu_{\ a} n^\nu \gamma^\sigma_{\ c} n^\rho C_{\mu\nu\sigma\rho}
= - \hat{R}_{ab} + K K_{ab} - K_{ac} K^{c}_{\ b}
+ \frac{2 \Lambda}{(d+1)} \gamma_{ab},            \label{pjwt}
\end{align}
where $\hat{R}_{abcd}$ and $\hat R_{ab}$ are the Riemann and Ricci
tensors of $\Sigma_c$.
Inserting \eqref{pjwt} into \eqref{pjbc} and inverting the relationship between
$K_{ab}$ and $t_{ab}$, the boundary condition \eqref{pjbc}
becomes
\begin{align}
& \frac{2}{\lambda^2} t^\tau_{\ i} t^\tau_{\ j}
+ \frac{t^2}{d^2} \gamma_{ij} + \frac{2 \Lambda}{d} \gamma_{ij}
- (t^\tau_{\ \tau} - 2 \lambda D_\tau)
\bigg(\frac{t}{d} \gamma_{ij}-t_{ij}\bigg)  \nonumber
\\& \qquad- \frac{2}{\lambda} D_{(i} t^\tau_{\ j)}
- t_{ik} t^k_{\ j}
- \hat{R}_{ij}=0.    \label{BYBC}
\end{align}
Note that the appearance of $\lambda$ in this equation comes purely from the
rescaling of the coordinate $x^0 \to \tau/\lambda$.

To see the effects of the fluctuation modes let us expand the metric, Ricci
tensor and the Brown-York tensor on $\Sigma_c$ in the following form,
\begin{align}
&\gamma_{ab} = \gamma_{ab}^{(B)}+\sum_{n=1}^\infty \gamma_{ab}^{(n)}\lambda^n,
\label{gamfluc} \\
&\hat R_{ab} = \hat R_{ab}^{(B)}
+ \sum^\infty_{n=1} \lambda^n \hat R_{ab}^{(n)},  \label{Rfluc}\\
&t^a_{\ b}= t^{a(B)}_{\ b} + \sum^\infty_{n=1} \lambda^n t^{a(n)}_{\ b},
\label{seby}
\end{align}
where the leading terms on the right hand side represent the background values.
In the near horizon limit, $f_c$ can be
rearranged in the form
\begin{align*}
&f_c = f'_h \cdot (\epsilon \,\alpha^2\lambda^2)
+\frac{1}{2}f''_h \cdot (\alpha^2\lambda^2)^2, \\
& f'_h =-\bigg(\omega+\frac{4\Lambda r_h}{d} \bigg),\quad
f''_h = - \frac{4\Lambda}{d}.
\end{align*}
which follows from the Taylor series expansion around $r_h$.
Consequently, the background values given in \eqref{tb} will also develop
$\lambda$ dependences. In the end, we have
\begin{align}
&t^\tau_{\ \tau} = 0 +\lambda t^{\tau(1)}_{\ \tau}
+ \cdots,   \nonumber\\
&t^\tau_{\ i}=0+\lambda t^{\tau(1)}_{\ i}+\cdots,  \nonumber\\
&t^i_{\ j} = \frac{\epsilon}{2}
\Big(\frac{\sqrt{\epsilon f'_h}}{\alpha \lambda}
- \alpha \lambda \frac{4\Lambda}{d\sqrt{\epsilon f'_h}}\Big)
\delta^i_{\ j} + \lambda t^{i(1)}_{\ j} + \cdots,  \nonumber\\
&t = \frac{d\epsilon}{2}
\Big(\frac{\sqrt{\epsilon f'_h}}{\alpha \lambda}
- \alpha \lambda \frac{4\Lambda}{d\sqrt{\epsilon f'_h}}\Big)
+ \lambda t^{(1)} + \cdots.
\label{PBY}
\end{align}

What we need to do is to expand \eqref{momc}, \eqref{hamc} and \eqref{BYBC} 
into power series
in $\lambda$ and look at the first nontrivial order contributions. For this
purpose we need to supplement \eqref{PBY} with expansions of $\hat R_{ab}$
and $D_{(i} t^\tau_{\ j)}$. To evaluate the latter we need to expand the
Christoffel connection $\hat \Gamma^a_{bc}$. Omitting the details we present
the results as follows:
\begin{align}
&\hat \Gamma^a_{bc}=\Gamma^{a(B)}_{bc} + \mathcal{O}(\lambda^1), \nonumber\\
&\hat R_{ab}=\hat R^{(B)}_{ab} + \mathcal{O}(\lambda^1),\label{backR}
\end{align}
where the background values
\begin{align*}
&\hat \Gamma^{\tau(B)}_{ab}=\hat \Gamma^{a(B)}_{\tau b}=0,
&&\hat \Gamma^{k(B)}_{ij} =\frac{1}{2}\left(\delta^k{}_i\partial_j\Phi
+\delta^k{}_j\partial_i\Phi - \delta_{ij}\partial^k \Phi\right),\\
&\hat{R}_{\tau a}^{(B)}=0,
&&\hat{R}_{ij}^{(B)}=\frac{2\Lambda}{d}e^\Phi \delta_{ij}
\end{align*}
are all $\lambda$-independent. Therefore, we have
\begin{align}
D_{(k} t^{\tau}_{\ j)} &= \partial_{(k}
\big(\lambda t^{\tau(1)}_{\ j)}
+ \mathcal{O}(\lambda^{2})\big)
- \big(\hat{\Gamma}^{l(B)}_{\ (kj)}
+ \mathcal{O}(\lambda^{1})\big)
\big(\lambda t^{\tau(1)}_{\ l}
+ \mathcal{O}(\lambda^{2})\big)    \nonumber\\
&\equiv \lambda \zeta_{kj}^{(1)} +
\lambda^2 \zeta_{kj}^{(2)} + \mathcal{O}(\lambda^{3}), \label{Dt}
\end{align}
where
\begin{align}
&\zeta_{kj}^{(1)}
= \partial_{(k} t^{\tau(1)}_{\ j)}
- \partial_{(k} \Phi t^{\tau(1)}_{\ j)}
+ \frac{1}{2} \delta_{kj} \delta^{lm}
\partial_l\Phi t^{\tau(1)}_{\ m},      \label{zeta}
\end{align}
and $\zeta_{kj}^{(2)}$ is some complicated expression which depends on 
$\gamma^{(1)}_{kj}$ and $t^{\tau(2)}_{\ j}$. We do not need to use the explicit 
form of $\zeta_{kj}^{(2)}$ in this paper.

Substituting eqs. \eqref{PBY} and \eqref{Dt}
into \eqref{BYBC}, we get in the first
nontrivial order $\mathcal{O}(\lambda^{0})$ the following identity,
\begin{align}
t^{i(1)}_{\ j}
&=\frac{\alpha\epsilon}{\sqrt{\epsilon f'_h}} \cdot
2\gamma^{ik(0)}
\big(t^{\tau(1)}_{\ k} t^{\tau(1)}_{\ j} - \zeta_{kj}^{(1)}\big)
+ \frac{1}{d} t^{(1)} \delta^{i}_{\ j},
\label{PBYBC}
\end{align}
where $\gamma^{ik(0)}= e^{-\Phi}\delta^{ik}$. Notice that the dependences on
$\alpha$ and $\epsilon$ can be get rid of by choosing $\alpha = \epsilon
\sqrt{\epsilon f'_c}$, which we will take from now on.
Similarly, expanding the $\tau$ component of \eqref{momc} into power series
in $\lambda$ yields, at order $\mathcal{O}(\lambda^{-1})$,
\begin{align}
\delta^{ij}\bigg(
\partial_i + \frac{d-2}{2} \partial_i\Phi
\bigg)  t^{\tau(1)}_{\ j}=0, \label{by1}
\end{align}
and expanding the spatial components of the same equation
yields, at the order $\lambda^{1}$, the following equation,
\begin{align}
\partial_\tau t^{\tau(1)}_{\ i}
-\frac{1}{2} \big(t^{(1)}-t^{\tau(1)}_{\ \tau} \big)\partial_i \Phi
+\bigg(\partial_j + \frac{d}{2}\partial_j \Phi\bigg) t^{j(1)}_{\ i}
= 0. \label{by2}
\end{align}
The lowest order contribution to the Hamiltonian constraint is at order
$\mathcal{O}(\lambda^0)$, which yields
\begin{align}
t^{\tau(1)}_{\ \tau} = -2 \gamma^{ij(0)}
t^{\tau(1)}_{\ i} t^{\tau(1)}_{\ j}. \label{hamc1}
\end{align}
The form of the equations \eqref{PBYBC}, \eqref{by1}, \eqref{by2}
and \eqref{hamc1} is exactly
the same as those appeared in \cite{HWZ} -- though with different values for
each quantity entering the equations -- which had led to a flat space
fluid system after introducing some appropriate holographic dictionary.
Therefore, we can follow the same line of argument as made in \cite{HWZ}
and introduce the holographic dictionary
\begin{align}
\rho = e^{\frac{d}{2}\Phi} , \quad
\mu = e^{\frac{d-2}{2}\Phi} ,\quad
\nu = \frac{\mu}{\rho} = e^{-\Phi} ,\quad
\end{align}
and
\begin{align}
\quad t^{\tau(1)}_{\ i} = \frac{v_i}{2\nu}  ,\quad
\quad \frac{t^{(1)}}{d}=\frac{p}{2\mu},
\end{align}
where $\rho, \mu, v_i, p$ are to be interpreted as the density, viscosity,
velocity field and the pressure of the dual fluid. Finally, eq. \eqref{by1}
becomes the continuity equation
\begin{align}
\partial^j (\rho  v_j) = 0 , \label{Cont}
\end{align}
and eqs. \eqref{PBYBC} and \eqref{by2} combined together give rise to the
standard equation
\begin{align}
\rho (\partial_\tau v_i + v^j \partial_j v_i)
= - \partial_i p + \partial^j d_{ij} + f_i
\label{NV}
\end{align}
for the velocity field of the fluid, where
\begin{align}
d_{ij} = \mu \bigg( \partial_j  v_i + \partial_i  v_j
- \frac{2}{d} \delta_{ij} \partial^k v_k \bigg) \label{dij}
\end{align}
represents the deviatoric stress, which is symmetric traceless and depends only
on the derivatives of the velocity field, hence vanishes in the hydrostatic
equilibrium limit, and
\begin{align}
f_i = \partial^j \Phi \bigg( d_{ij} + \frac{d-2}{2} p \delta_{ij}\bigg)
+ \frac{2}{d} v^j v_j \partial_i \rho
- \frac{2}{d} (v^j \partial_j \rho) v_i   \label{bdyfc}
\end{align}
represents an extra force. Since the first two terms $- \partial_i p
+ \partial^j d_{ij}$ on the right hand side of \eqref{NV} correspond to
the ordinary surface forces, the last term which cannot be cast in the
form of the first two terms must represent a body force. The equations
\eqref{Cont} and \eqref{NV} constitute a system of equations
governing the motion of a compressible, forced, stationary and viscous fluid
moving in the $(d+1)$-dimensional Newtonian spacetime
$\mathbb{R}\times\mathbb{E}^d$.

\section{Finite cutoff}

The whole construction of the last section is very similar to that made in
\cite{HWZ} for black hole background. Now we would like to ask a different
question: Can we construct a flat space fluid dual without making use of the
near horizon Petrov I boundary condition?

To answer this question, let us go over the whole process of the construction.
It is clear the every formulae until \eqref{seby} still holds if $\Sigma_c$
is not placed near the horizon. However, from eq. \eqref{PBY} and onwards,
things begin to change a little. Concretely, \eqref{PBY} must be replaced by
\begin{align}
&t^\tau_{\ \tau} = 0 +\lambda t^{\tau(1)}_{\ \tau}
+ \cdots,   \nonumber\\
&t^\tau_{\ i}=0+\lambda t^{\tau(1)}_{\ i}+\cdots,    \nonumber\\
&t^i_{\ j} =\frac{f^\prime_c}{2\sqrt{f_c}}\delta^i_{\ j}
+\lambda t^{i(1)}_{\ j} + \cdots,     \nonumber\\
&t=\frac{d}{2} \frac{f^\prime_c}{\sqrt{f_c}}
+ \lambda t^{(1)} + \cdots,
\label{PBY1}
\end{align}
where $f_c, f'_c$ etc must be kept un-expanded. If we replace \eqref{PBY}
with \eqref{PBY1} and carry on the rest process as made in the last section,
then we will encounter some trouble. Unlike what has been derived in the
near horizon limit, at finite cutoff the first nontrivial order of the
boundary condition becomes
\begin{align}
t^{\tau(1)}_{\ k} t^{\tau(1)}_{\ j} - \zeta^{(1)}_{kj} = 0.  \label{byo0}
\end{align}
insert \eqref{byo0} into \eqref{by1}, we could only come
to the conclusion that at order $\mathcal{O}(\lambda^{1})$ the
momentum constraints will not allow fluctuation in the
$t^\tau_{\ i}$ component, i.e.
\begin{align}
t^{\tau(1)}_{\ i} = 0.
\end{align}
If we expand the boundary condition to the next
order, then the following relation would follow,
\begin{align}
t^{i(1)}_{\ j} = \frac{t^{(1)}}{d} \delta^i_{\ j}
- \frac{\sqrt{f_c}}{f'_c}\big(R^{i(1)}_{\ j}
+ 2\zeta^{i(2)}_{\ j}\big), \label{z2}
\end{align}
where
\[
\zeta^{i(2)}_{\ j}= \gamma^{ik(1)}\zeta^{(1)}_{kj}
+\gamma^{ik(0)}\zeta^{(2)}_{kj}.
\]
Using \eqref{z2} we can the spatial component of
the momentum constraints \eqref{by2} into the following equation:
\begin{align}
\frac{1}{d} \partial_i t^{(1)}
&= \frac{\sqrt{f_c}}{f'_c}
\bigg(\partial_j + \frac{d}{2} \partial_j \Phi\bigg)
\Big(R^{j(1)}_{\ i} + 2\zeta^{j(2)}_{\ i}\Big) \nonumber\\
&= \frac{\sqrt{f_c}}{f'_c} e^{-\frac{d}{2}\Phi}\partial_j
\Big[e^{\frac{d}{2}\Phi} \big(R^{j(1)}_{\ i}
+ 2\zeta^{j(2)}_{\ i}\big)\Big],  \label{byo1}
\end{align}
but it is exceedingly hard to explain \eqref{byo1}
as a partial differential equation that describes
the fluid motion.

To avoid the above problem and realize the fluid dual at finite
cutoff, we employ a translation of $\Phi$ as follows:
\begin{align}
\Phi \rightarrow \Phi - \ln \lambda.  \label{transf}
\end{align}
Such a translation does not alter the order of derivatives of $\Phi$ but
does change the order of $e^\Phi$. 
After the translation \eqref{transf}, the spatial components
of the induced metric \eqref{dsd} will be rescaled:
\begin{align}
\gamma^{(B)}_{ab} \mathrm{d} x^a \mathrm{d} x^b \rightarrow
\tilde{\gamma}^{(B)}_{ab} \mathrm{d} x^a \mathrm{d} x^b
= -\frac{1}{\lambda^2} \mathrm{d}\tau^2
+ \frac{e^\Phi}{\lambda} \delta_{ij}
\mathrm{d} x^i \mathrm{d} x^j,
\end{align}
and eq. \eqref{GLveq} becomes
\begin{align}
& \delta^{jk} \partial_j \partial_k \Phi
+ \frac{d-2}{2}\bigg( 2 \partial_i^2 \Phi
+ \delta^{jk} \partial_j \Phi \partial_k \Phi
- (\partial_i \Phi)^2  \bigg)
+ \frac{4\Lambda}{d} \frac{e^\Phi}{\lambda}=0.
\end{align}
As a result the Ricci tensor of $\Sigma_c$ will be inversely proportional to
$\lambda$,
\begin{align}
\hat{R}_{ij}^{(B)}
\rightarrow \tilde{R}_{ij}^{(B)}
=\frac{2\Lambda}{d}
\frac{e^\Phi}{\lambda} \delta_{ij}.
\end{align}
The background Brown-York tensor
$t^{a(B)}_{\ b}$, the Christoffel connection
$\hat{\Gamma}^{a(B)}_{\ bc}$ and $\zeta^{(1)}_{ij}$ are all kept invariant.
So we can list the new boundary condition as
\begin{align}
& \frac{2}{\lambda^2} \tilde{\gamma}^{ik} t^\tau_{\ k} t^\tau_{\ j}
+ \frac{t^2}{d^2} \delta^i_{\ j} + \frac{2 \Lambda}{d} \delta^i_{\ j}
- (t^\tau_{\ \tau} - 2 \lambda D_\tau)
\bigg(\frac{t}{d} \delta^i_{\ j} - t^i_{\ j}\bigg)  \nonumber\\
& \qquad - \frac{2}{\lambda} \tilde{\gamma}^{ik} D_{(k} t^\tau_{\ j)}
- t^i_{\ k} t^k_{\ j}
- \hat{R}^i_{\ j}=0. \label{BYBC1}
\end{align}
The form of this equation looks identical to \eqref{BYBC}, however with
a different background metric $\tilde{\gamma}^{ij}$ which is $\lambda$
dependent. Using \eqref{BYBC1} in place of \eqref{BYBC} and 
inserting \eqref{Dt} and \eqref{PBY1}, we get the following new equation at the 
order $\mathcal{O}(\lambda^1)$,
\begin{align}
t^{i(1)}_{\ j}
=\frac{\sqrt{f_c}}{f^\prime_c} \cdot
2 {\gamma}^{ik(0)}
\big(t^{\tau(1)}_{\ k}
t^{\tau(1)}_{\ j}
- \zeta^{(1)}_{kj}\big)
+ \frac{1}{d} t^{(1)} \delta^{i}_{\ j}
- \frac{\sqrt{f_c}}{f^\prime_c} \hat{R}^{i(1)}_{\ j}.
 \label{BYBCfc}
\end{align}

Obviously the momentum constraints are 
invariant under the translation \eqref{transf} for $\Phi$,
so there will not be any new
terms in equation \eqref{by1} and \eqref{by2}. 
Substituting \eqref{BYBCfc} into \eqref{by2}, and introducing
the following holographic dictionary
\begin{align}
& \rho = \frac{f^\prime_c}{\sqrt{f_c}} e^{\frac{d}{2}\Phi}, \quad
\mu = e^{\frac{d-2}{2}\Phi} ,\quad
\nu = \frac{\mu}{\rho}
= \frac{\sqrt{f_c}}{f^\prime_c} e^{-\Phi} ,  \nonumber\\
& t^{\tau(1)}_{\ i} = \frac{v_i}{2\nu}  ,\quad
\quad \frac{t^{(1)}}{d}=\frac{p}{2\mu},
\end{align}
eqs. \eqref{by1} and \eqref{by2} will again become
the continuity equation and the Navier-Stokes
equation
\begin{align}
&\partial^j (\rho  v_j) = 0 ,   \label{Cont1}\\
&\rho (\partial_\tau v_i + v^j \partial_j v_i)
= - \partial_i p + \partial^j d_{ij} + f_i,   \label{NV1}
\end{align}
where $d_{ij}$ takes the same form as in \eqref{dij} (of course with different 
$\mu$ and $v_i$), and
\begin{align}
f_i = \partial^j \Phi \bigg( d_{ij} + \frac{d-2}{2} p \delta_{ij}\bigg)
+ \frac{2}{d} v^j v_j \partial_i \rho
- \frac{2}{d} (v^j \partial_j \rho) v_i
+ \frac{2\sqrt{f_c}}{f^\prime_c}
\nu \partial_j (\rho \hat{R}^{j(1)}_{\ i}).   \label{bdyfc2}
\end{align}
Comparing to the case of the last section, we see that an extra force term
$\frac{2\sqrt{f_c}}{f^\prime_c}
\nu \partial_j (\rho \hat{R}^{j(1)}_{\ i})$ appears in $f_i$ at finite cutoff.

\section{Concluding remarks}

The result of this work indicates that a compressible, forced, viscous fluid
in flat Newtonian spacetime can not only be realized as the holographic dual of 
fluctuation modes around a black hole background, but also be realized as the 
dual of fluctuating modes around a product Einstein manifold. The construction
relies on taking a timelike hypersurface $\Sigma_c$ 
-- which is placed either near 
the horizon or at some finite cutoff -- 
and mapping it into an Euclidean 
space which lies in its conformal class. It is surprising that the final
fluid equations, including the form of the extra force term $f_i$, are 
basically the same as that arising from fluctuations around a black hole 
background (though in the case of a finite cutoff a novel force term appears). 
It remains to understand the nature of the extra force term from the point 
of view of fluid dynamics.

Experienced readers on Gravity/Fluid correspondence might feel that
the fluid equations in flat space are only those for a curved space 
incompressible fluid in disguise. This is true in some sense. However, 
writing the fluid equations in flat Newtonian spacetime is still a nontrivial 
step forward, because the holographic dictionary we adopt is quite different for 
that leading to a curved space incompressible fluid. 
In practice, any physical system which is dual to another one in some way
can be regarded as its dual in disguise. However this view point does not 
disvalue the duality relationship between the two systems.

As a technical addendum let us mention why we stick to the approach which
makes use of the Petrov I boundary condition rather than adopting 
the boost-rescaling approach which is also widely used in Gravity/Fluid
correspondence \cite{Compere,cai1,Eling,Compere:2012mt,
Eling:2011cl,Bai:2012ci,Zou:2013ix,
Hu:2013dza,Niu:2011gu,Cai:2012mg,Cai,1204.2029}. 
The reason lies in that the background geometry which we use
is not necessarily flat, and boosting requires the existence of a flat boundary.
For this reason, we do not expect the boost-rescaling approach to be
applicable in our situations.

The product manifold taken in this work is of the simplest kind, i.e. the total 
manifold is the product of two submanifolds. We can of course consider the case
when the total manifold is the product of several submanifolds, e.g. 
the submanifold $\mathcal{M}_2$ itself has a product structure, with
line element of the form 
\[
\mathrm{d}s_{\mathcal{M}_2}^2 
= e^{\Phi(x)}\delta_{ij}\mathrm{d}x^i \mathrm{d}x^j
+e^{\Psi(y)}\delta_{mn}\mathrm{d}y^m \mathrm{d}y^n+\cdots.
\]
We expect that flat space fluid may still be constructed out of the fluctuating 
modes around such backgrounds, and the resulting fluid might become  
anisotropic. We will have more to say along the lines of 
research of \cite{HWZ} and the present work.

\section*{Appendix}

In this appendix we discuss some properties regarding the system of equations
\eqref{GLveq} and \eqref{GLveq2} that determine the function $\Phi(x)$.

For the particular case of $d=2$, these equations degenerate
into a single equation
\[
\delta^{jk} \partial_j \partial_k \Phi + 2\Lambda e^\Phi=0,
\]
which can be easily recognized to be the Euclideanized Liouville equation
for $\Lambda\neq 0$ or the Laplacian equation for $\Lambda=0$. The fact
that the metric function $\Phi$ obeys Liouville or Laplacian equation in the 
case $d=2$ is first observed in \cite{Moskalets:2014hoa} while considering 
black hole metrics. In the case of product Einstein manifold such equations 
still hold. Existence of infinitely many solutions to such equations are well 
known.

For $d>2$, these equations are much more complicated. However, there is an
explicit solution which we reproduce below. Let 
\begin{align}
\Phi=- 2\log \chi.\label{phichi}
\end{align}
Then, eq. \eqref{GLveq2} becomes
\[
\partial_i \partial_j \chi =0, \quad i\neq j.
\]
Therefore, $\chi$ must take the form
\[
\chi = \sum_i \chi_i(x^i),
\]
where each term depends on a single $x^i$. Moreover, eq. \eqref{GLveq} can
be rewritten as
\begin{align}
\sum_k \frac{(d-1)(\partial_k \chi)^2 -\chi \partial_k^2 \chi}{\chi^2}
-(d-2)\frac{\chi \partial_i^2 \chi}{\chi^2}
+\frac{4\Lambda}{d\chi^2} =0. \label{eqab}
\end{align}
Suppose
\[
\chi = a+ b\sum_i x_i^2,
\]
eq. \eqref{eqab} becomes an algebraic equation
\begin{align*}
\frac{\Lambda}{d} - a b (d-1)=0
\end{align*}
for the coefficients $a$ and $b$. When $\Lambda\neq 0$, both $a$ and $b$ 
cannot vanish. So, we can freely choose $a=1$ 
and then get
\[
b= \frac{\Lambda}{d(d-1)}.
\]
This yields the solution
\[
\chi(x) = 1+ \frac{\Lambda}{d(d-1)}\sum_k x_k^2,
\]
i.e.
\[
e^\Phi = \frac{1}{\chi^2}
= \frac{1}{\bigg(1+ \frac{\Lambda}{d(d-1)}\sum_k x_k^2\bigg)^2}.
\]
If $\Lambda=0$, then either $a$ or $b$ must be zero. Taking $a=0$ and let $b$ 
be arbitrary, we get
\[
\chi(x) = b\sum_k x_k^2,
\]
i.e.
\[
e^\Phi = \frac{1}{\chi^2}
= \frac{1}{b^2\bigg(\sum_k x_k^2\bigg)^2}.
\]
$e^\Phi$ becomes singular in this case at $x_i=0$. Alternatively, we can 
take $b=0$ and normalize $a=1$, which corresponds to $\chi=1$, i.e. $\Phi=0$. 
In this particular case the fluid we obtain becomes 
{\em incompressible} but still
lives in flat space, and the force $f_i$ given in \eqref{bdyfc} (and also the 
first two terms in \eqref{bdyfc2}) becomes vanishing. Since eqs. \eqref{GLveq} 
and \eqref{GLveq2} are essentially nonlinear, there may be other solutions to 
these equations.

\providecommand{\href}[2]{#2}\begingroup
\footnotesize\itemsep=0pt
\providecommand{\eprint}[2][]{\href{http://arxiv.org/abs/#2}{arXiv:#2}}


\end{document}